\documentstyle[twoside,fleqn,espcrc2,epsfig]{article}

\newcommand{\AmS}{{\protect\the\textfont2
  A\kern-.1667em\lower.5ex\hbox{M}\kern-.125emS}}

\renewcommand{\Re}{\mathop{\rm Re}\nolimits}

\newcommand{\tr}{\mathop{\mathrm{tr}}}

\title{The U(1) phase transition on toroidal and spherical lattices}

\author{I. Campos\thanks{Speaker at the conference}, 
A. Cruz and A. Taranc\'on \\
{Departamento de
F\'{\i}sica Te\'orica, Universidad de Zaragoza, \\ 
        C/ Pedro Cerbuna 12, E-50009 Zaragoza. Spain}%
        }
       
\begin{document}

\begin{abstract}
We have studied the properties of the phase transition in the U(1)
compact pure gauge model paying special atention to the influence 
of the topology of the boundary conditions.
From the behavior of the energy cumulants and the observation
of an effective $\nu \sim 1/d$ on toroidal and spherical lattices,
we conclude that the transition is first order.

\end{abstract}

\maketitle


We have focussed on the problem of the influence
of the topology of the boundary conditions on the properties of the
phase transition in the compact pure gauge U(1) model.
Its most popular lattice formulation is obtained through
the Wilson action:
\begin{equation}
S_{\rm W} = \beta \sum_{\rm P} [1 - \Re \tr U_{\rm P}] 
\label{SW} 
\end{equation}
for which most of the numerical work has been done. Lattices as large
as $L=16$ have been simulated \cite{ALFON}, finding two-state
distributions from $L=6$ on, increasing free energy gaps, and a $\nu$
exponent compatible with $1/d$ has been observed as well \cite{ALFON}.
Altogether the transition is commonly believed to be first order,
and since $\xi_c$ remains finite, no consequences for possible continuum limits
should be expected.

Enlarging the parameter space by adding a term to the action
in the adjoint representation
\begin{equation}
S_{\rm {EW}} = S_{\rm W} +
\gamma \sum_{\rm p} [1 - \Re \tr U_{\rm p}^2]
\label{SEW} 
\end{equation}
does not seem to change qualitatively the first order
scenario \cite{SEGUNDO}. 

Mainly two issues remain still to be clarified in order to definitively
discard the possibility of, despite all numerical evidence, having
a continuous transition in the thermodynamical limit:
1) On the one hand, a complete stabilization of the latent heat
has not been observed in numerical simulations, and hence
the possibility of a thermodynamical limit where the observed two peaks 
superimpose is still open;
2) on the other hand, the role of non trivial topological structures
appearing on finite toroidal lattices is a source of controversy since
wrapping monopoles were conjectured to be responsible for the energy 
jump. Using in the simulations lattices homotopic to the sphere allows
monopole loops to be contracted to a single point.
Should this hypothesis be correct, no energy jump will be observed
in the simulations on spherical lattices. Indeed this is the behavior
observed in \cite{CN,BAIG,JCN}. However, the spherical lattices constructed
in those simulations are not homogeneous, and presumably, larger lattices 
should be needed to get rid of uncontrolled finite size effects.

We have performed a comparative study on the toroidal and spherical 
topologies to shed some light on both problems \cite{SEGUNDO}.
The spherical lattice is constructed by considering the surface of a $5D$ cube 
which is topologically equivalent to a $4D$ sphere \cite{CN,JCN}.
On this $4D$ surface there is a number of sites with less than eight 
surrounding links, the homogeneity being only restored in the 
thermodynamical limit. To alleviate these
inhomogeneities authors in \cite{JCN} increase the contribution to the action 
of the inhomogeneous sites by an amount proportional to its lack of neighbors.
We do not expect this smoothing to affect the existence of two states
nor the Finite Size Scaling properties in large enough lattices.


We use the extended Wilson action (\ref{SEW}) and define 
the plaquette energy in the usual way,
$ E_{\rm p} = \frac{1}{N_{\rm p}} \langle \sum_{\rm p} \cos \theta_{\rm p} 
\rangle $, where $N_p$ denotes the number of plaquettes.
On the torus $N_p = 6 L^4$; on the spherical
lattice $N_p$ is not simply proportional to the number of sites, but it
can be computed as a function of the base length, $N$, of the
$5D$ cube, $N_p = 60(N-1)^4+20(N-1)^2$.

We tipically perform trial runs to locate the
peak of the specific heat, $\beta^{\ast}(L)$, where
we perform a simulation to get the energy distribution,
$P_E(\beta^{\ast})_L$.

On the torus we worked at $\gamma =-0.1, -0.2$,  
in lattice sizes up to $L=20$ 
and up to $L=24$ at $\gamma=-0.3,-0.4$; 
on the sphere we first study the Wilson action ($\gamma=0$) to
check for the absence of two state signals claimed by authors in 
\cite{CN,JCN}. To compare with our results on the torus
we also simulate at $\gamma = -0.2$.

\begin{figure}[htb]
\epsfig{figure=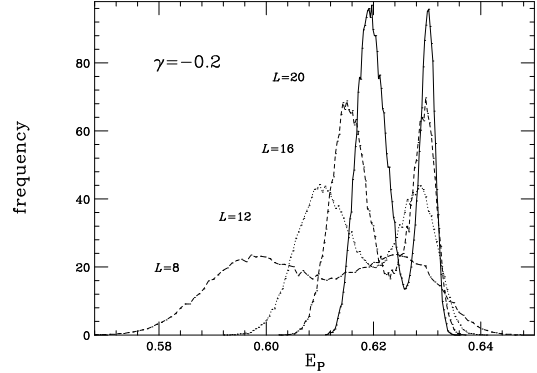,angle=90,width=70mm}
\caption{$E_p$ histograms on the torus at $\gamma=-0.2$.}
\label{HT}
\end{figure}

From our simulations on the torus a two state signal together
with an increasing free energy gap is revealed by
the histograms for all lattice sizes, at all $\gamma$ values we simulate
(see Figure \ref{HT}).
The lattice size at which the free energy gap starts appearing is larger
the more negative $\gamma$ is, and the latent heat decreases as $\gamma$
gets more negative.

\begin{figure}[htb]
\epsfig{figure=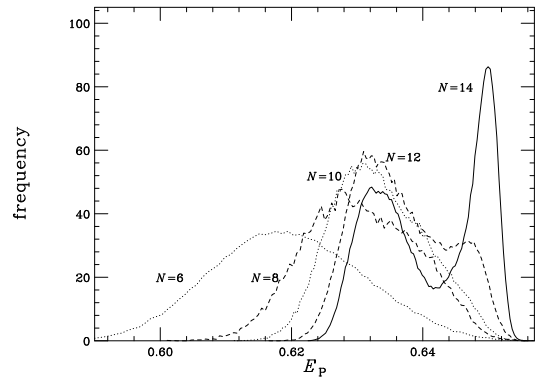,angle=90,width=70mm}
\caption{$E_p$ histograms on the sphere at $\gamma=0$.}
\label{HS}
\end{figure}

On the spherical lattice, at $\gamma=0$ we find two state signals
from $N=12$ on. An increasing energy gap is observed when simulating $N=14$,
(see Figure \ref{HS}) together with a volume increasing rate in the
specific heat maximum. At $\gamma = -0.2$ the two state signal sets in at
$N=16$. Comparing with the toroidal lattices for which equivalent signals
are observed ($L=6$ at $\gamma=0$; $L=12$ at $\gamma = -0.2$) a first 
observation is the retard on the onset of double peak distributions
on the sphere by a factor around 100 in volume.

From the energy distributions we compute the position of the  
nearby partition function zero closest to the real axis,
whose scaling law allows to compute
the $\nu$ exponent: $Im(\omega_0) \sim L^{-1/\nu}$.
In order to monitorize the phase transition with increasing lattice size
an effective exponent, $\nu_{\rm eff}$, is computed following:
\begin{equation}
\nu_{\rm eff} = - \frac{\ln L_2/L_1}{ \ln (Im\omega_0(L_2)/Im\omega_0(L_1))}
\end{equation}

\begin{figure}[htb]
\epsfig{figure=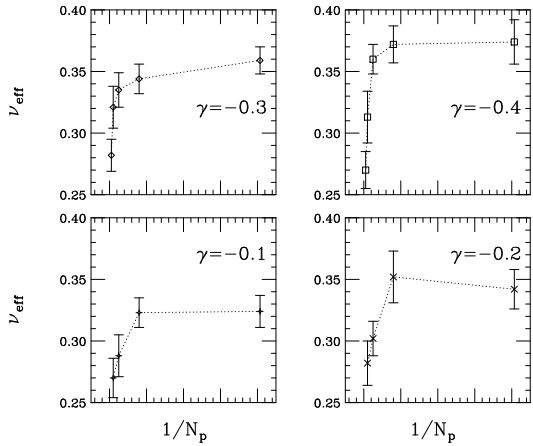,angle=90,width=70mm}
\caption{Exponent $\nu_{\rm {eff}}$ on the torus.}
\label{NUTORO}
\end{figure}

On the torus (see Figure \ref{NUTORO}) the quasi-stabilization of the latent
heat coincides with the falling of $\nu_{\rm {eff}}$ from a value around 1/3
towards the first order value $1/d$. This is the typical behavior 
expected for the effective exponents  
in weak first order phase transitions \cite{MARIA}.


From the energy jump on finite lattices, $C_{\rm {lat}}(L)$,
we compute the latent
heat on the thermodynamical limit, $C_{\rm {lat}}(\infty)$, on the torus. 
On a lattice with periodic
boundary conditions the FSS behavior of a $\xi$ dependent
quantity, such as the latent heat,
is expected to be controlled by the $\nu$ exponent with a law:
\begin{equation}
C_{\rm {lat}} (L) = C_{\rm {lat}}(\infty) + A L^{-1/\nu}
\label{EX}
\end{equation}
We get a $C_{\rm {lat}}(\infty) \neq 0$ for all $\gamma$ we investigate.

\begin{figure}[htb]
\epsfig{figure=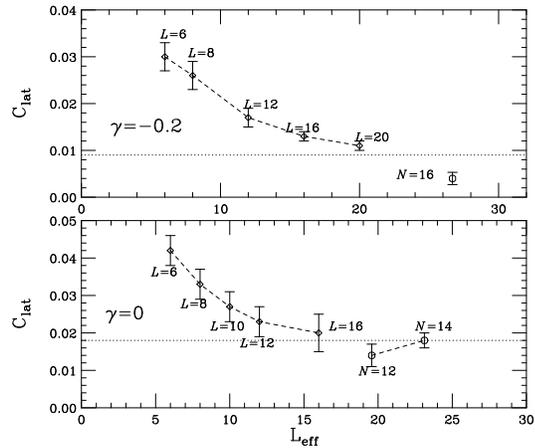,angle=90,width=70mm}
\caption{Latent Heat on the torus and on the sphere (circles).
The dotted line indicates $C_{\rm {lat}}(\infty)$ from (\ref{EX}).}
\label{LAT}
\end{figure}

On the sphere the behavior is not simple due to the inhomogeneities.
At $\gamma=0$ a shifting of the peak in the Coulomb phase is observed.
The explanation for this fact is the
smaller contribution to the action of the sites with less than maximum
connectivity. In the low $T$ region (Coulomb phase) their influence is
stronger since the system is more ordered. As larger lattices are considered
those sites contributions are less and less important, and the
distance between the two peaks tends to the latent heat obtained by
extrapolating the results obtained in the torus (see Figure \ref{LAT}).
The conjectures about a possible superimposition of the two peaks
in the thermodynamical limit becomes rather unplausible when the results
on spherical lattices are taken into account, since both, the sphere and the
torus share a common behavior in this limit.

To summarize, our results show a clear first order behavior in all the 
observables we have studied. The observation of an energy jump in the
spherical lattice seems to rule out the conjectures about the influence
of non trivial monopoles on such jump. It is worth remarking that  
simulations suppresing wrapping monopoles by other techniques
were performed some time ago \cite{LIPPERT}, 
their results showing the existence of an energy jump as well.

\end{document}